\documentclass[twocolumn,showpacs,preprintnumbers,amsmath,amssymb]{revtex4}
\usepackage{graphicx}
\usepackage{dcolumn}
\usepackage{bm}
\usepackage{subfigure}
\usepackage{color}

\begin{document}

\title{Possible indication to the QCD evolution of double parton distributions?}


\author{A.M.~Snigirev}
\affiliation{M.V. Lomonosov Moscow State University, D.V. Skobeltsyn
Institute of Nuclear Physics, 119991, Moscow, Russia }

\date{\today}
\begin{abstract}
For the first time the process-independent parameter of double parton scattering, $\sigma_{\rm eff}^{\rm exp}$, has been measured newly in the D0 experiment at the three different resolution scales. If we interpret the measurement as a decrease of the effective cross section with a growth of the resolution scale it can indicate the QCD evolution of double parton distributions.
\end{abstract}


\maketitle
\section{\label{sec1}Introduction}
The presence of multiple parton interactions in high energy hadron collisions has been convincingly  demonstrated by the AFS~\cite{AFS}, UA2~\cite{UA2} and CDF~\cite{cdf_4jets} Collaborations using events with  the four-jet final state and later by the CDF Collaboration~\cite{cdf} using events with the $\gamma+3$ jets final state. Recently the D0 Collaboration has measured the process-independent parameter of double parton scattering, $\sigma_{\rm eff}^{\rm exp}$, using the $\gamma+3$ jets events at the three different resolution (energy) scales~\cite{D0} providing new and complementary information on the proton structure. The possibility of observing two separate hard  collisions has been proposed since long~\cite{landshoff}, and from that has also developed in a number of works~\cite{takagi, trelani, sjostrand2, sjostrand3, maina, strikman}. A brief review of the current situation and some progress in the modeling with account of correlated flavor, color, longitudinal and transverse momentum distributions can be found in Ref.~\cite{sjostrand2}. 

Multiple interactions require an ansatz for the structure of the interacting hadrons, i.e. correlations between the constituent partons. As a simple ansatz, usually,  the two-parton distributions (of the parton momentum fraction) are supposed to be the product of two single-parton distributions times a momentum conserving phase space factor. In recent papers~\cite{snig03} it has been shown that this hypothesis is in some contradiction with the leading logarithm approximation of perturbative QCD  (in the framework of which a parton model, as a matter of fact, was established in the quantum field theories~\cite{gribov}). Namely, the two-parton distribution functions being the product of  two single distributions at some reference resolution scale become dynamically correlated at any different scale of a hard process. 

The main purpose of the present letter is to show that these QCD dynamical correlations result effectively in a dependence of the experimentally {\it measured} effective cross section $\sigma_{\rm eff}^{\rm exp}$ on the resolution scale unlike naive accepted expectations. 

The paper is organized as follows. Section~\ref{sec2}  is devoted to what is known from perturbative QCD theory on the two-parton distribution functions. The possible manifestation of correlations induced by QCD evolution is discussed in Sec.~\ref{sec3}. We summarize and conclude in Sec.~\ref{sec4}.

\maketitle
\section{\label{sec2} Double parton distributions in the leading logarithm approximation}

In order to introduce the denotations and to be clear  let us recall that, for instance, 
the inclusive differential cross section for the four-jet production due to the simultaneous interaction of two parton pairs within one $p{\bar p}$ collision, $\sigma_{\rm dp}$, can be given by~\cite{takagi}  
\begin{equation} 
\label{fourjet}
\sigma_{\rm dp} = \sum \limits_{q/g} \int \frac{ \sigma_{12} \sigma_{34}}
{2 \sigma_{\rm eff}} D_ p(x_1,x_3)D_{\bar{p}}(x_2,x_4)dx_1dx_2dx_3dx_4, 
\end{equation} 
where $\sigma_{ij}$ stands for the two-jet production cross section. The dimensional
phenomenological parameter $\sigma_{\rm{eff}}$ in the denominator  is a factor characterizing a size of the effective interaction region of the hadron (the factor 2 is introduced due to the identity of the two parton subprocesses). The relatively small value of $(\sigma_{\rm eff})_{\rm CDF}$ measured by the CDF Collaboration~\cite{cdf} with respect to the naive expectation was, in fact, considered~\cite{trelani, Treleani_2007} as evidence of nontrivial correlation effects between partons in the transverse space. But, apart from these correlations, the longitudinal momentum correlations can also exist and 
they were investigated in Ref.~\cite{snig03}. The factorization ansatz is just applied to the two-parton distributions entering Eq.~(\ref{fourjet}):
\begin{equation} 
\label{factoriz}
D_ p(x_i,x_j, Q^2) = D_ p(x_i,Q^2)D_ p(x_j,Q^2)(1-x_i-x_j),
\end{equation} 
where  $D_ p(x_i,Q^2)$ are the single quark/gluon momentum distributions at the scale $Q^2$ (deter\-mi\-ned by a hard process).

However multi-parton distribution functions satisfy the generalized 
Gribov-Lipatov-Altarelli-Parisi-Dokshitzer (GLAPD) evolution equations derived for the first time in Refs~\cite{kirschner,snig} as well as single parton distributions satisfy more known and cited GLAPD equations~\cite{gribov,altarelli}. Under certain initial conditions these generalized equations lead to the solutions, which are identical with the jet calculus rules proposed originally for multiparton fragmentation functions by
Konishi-Ukawa-Veneziano~\cite{konishi} and are in some contradiction with the
factorization hypothesis (\ref{factoriz}). Here one should note that at the parton level this  is a strict assertion within the leading logarithm approximation. 

After introducing the natural dimensionless variable
$$t = \frac{1}{2\pi b} \ln \Bigg[1 + \frac{g^2(\mu^2)}{4\pi}b
\ln\Bigg(\frac{Q^2}{\mu^2}\Bigg)\Bigg]=\frac{1}{2\pi b}\ln\Bigg
[\frac{\ln(\frac{Q^2}{\Lambda^2_{\rm QCD} })}
{\ln(\frac{\mu^2}{\Lambda^2_{\rm QCD}})}\Bigg],$$
where $b = (33-2n_f)/12\pi~~
{\rm {in~ QCD}},$
$g(\mu^2)$ is the running coupling constant at the reference scale $\mu^2$, $n_f$ is the number of active flavors, $\Lambda_{\rm QCD}$ is the dimensional QCD parameter, the GLAPD equations read~\cite{gribov,altarelli}
\begin{equation}
\label{e1singl}
 \frac{dD_i^j(x,t)}{dt} = 
\sum\limits_{j{'}} \int \limits_x^1
\frac{dx{'}}{x{'}}D_i^{j{'}}(x{'},t)P_{j{'}\to j}\Bigg(\frac{x}{x{'}}\Bigg).
\end{equation}
\noindent
They describe the scaling violation of the parton distributions $ D^j_i(x,t)$ inside a dressed quark or gluon ($i,j = q/g$).

We will not write  the kernels $P$ explicitly and will not derive the generalized 
equations for  two-parton distributions $D_i^{j_1j_2}(x_1,x_2,t)$, representing the probability that in a dressed constituent $i$ one finds two bare partons  of  types $j_1$ and $j_2$ with the given longitudinal momentum fractions $x_1$ and $x_2$  (referring to~\cite{snig03,gribov,kirschner, snig, altarelli} for details). We note only that their solutions can be represented as the convolution of single distributions~\cite{kirschner, snig}. This convolution coincides with the jet calculus rules~\cite{konishi} as mentioned above and is the  generalization of the  well-known Gribov-Lipatov relation installed for single functions~\cite{gribov} (the distribution of bare partons inside a dressed constituent  is identical to the distribution of dressed constituents in the fragmentation of a bare parton in the leading logarithm approximation). The obtained solution  shows also that the double distribution of partons is {\it {correlated}} in the leading logarithm approximation:
\begin{eqnarray}
\label{nonfact}
D_i^{j_1j_2}(x_1,x_2,t) \neq D_{i}^{j_1}(x_1,t) 
D_{i}^{j_2}(x_2,t).
\end{eqnarray}

Of course, it is interesting to find out the phenomenological issue of this parton level consideration. This can be done within the well-known factorization of soft and hard stages (physics of short and long distances). As a result, the equations (\ref{e1singl}) 
describe the evolution of parton distributions in a hadron ($h$) with $t ~(Q^2)$, if one replaces the index $i$ by index $h$ only. However, the initial conditions for new equations at $t=0 ~(Q^2=\mu^2)$ are unknown {\it a priori} and must be introduced phenomenologically or must be extracted from experiments or some models dealing with physics of long distances [at the parton level: 
$D_{i}^{j}(x,t=0)~= ~\delta_{ij} \delta(x-1)$; ~$D_i^{j_1j_2}(x_1,x_2,t=0)~=~0$].
Nevertheless the solution of the generalized GLAPD evolution equations with a given initial
condition may be written as before via the convolution of single distributions~\cite{snig03,snig}. This result shows that if the two-parton distributions are factorized at some scale $\mu^2$, then the evolution violates this factorization {\it{ inevitably}} at any different scale ($Q^2 \neq \mu^2$), apart from the violation due to 
the kinematic correlations induced by the momentum conservation, which is the analogue of
the  momentum conserving phase space factor in Eq.~(\ref{factoriz}).

For a practical employment it is interesting to know the degree of this violation. Partialy this problem was investigated theoretically in Refs.~\cite{snig, snig2} and  for the two-particle correlations of fragmentation functions in Ref.~\cite{puhala}. That technique is based on the Mellin transformation of distribution functions and the asymptotic behavior can be estimated. Namely, with the growth of $t~(Q^2)$ the correlation term  becomes {\it {dominant}} for  finite $x_1$ and $x_2$~\cite{snig2} and thus the two-parton distribution functions ``forget'' the initial conditions unknown 
{\it a priori} and the perturbatively calculated correlations appear.

The asymptotic prediction ``teaches'' us a tendency only and tells nothing about the 
values of $x_1,x_2, t(Q^2)$ beginning from which the correlations are significant. Naturally numerical estimations can give an answer to this specific question using the CTEQ fit~\cite{cteq} for single distributions as an input. The nonperturbative initial conditions $D_h^j(x,0)$ are specified in a parametrized form at a fixed low-energy scale $Q_0=\mu=1.3$ GeV. The particular function forms and the value of $Q_0$ are not crucial for the CTEQ global analysis at a flexible enough parametrization. The results of numerical calculations were obtained in Ref.~\cite{snig03} for the ratio:
\begin{eqnarray}
\label{ratio}
R(x,t)~=
\frac{D_{p({\rm QCD,corr.})}^{gg}(x_1,x_2,t)}{ D_p^{g}(x_1,t)D_p^{g}(x_2,t)(1-x_1-x_2)^{2}} \Big|_{x_1=x_2=x}
\end{eqnarray}

At the hard process scale of the CDF measurement~\cite{cdf} ($Q \sim 6$ GeV) the ratio (\ref{ratio}) is nearly 10$\%$  and increases right up to 30$\%$ at much higher scale ($Q \sim 100$ GeV) for the longitudinal momentum fractions $x \leq 0.1$ accessible to these measurements. For the finite longitudinal momentum fractions $x \sim 0.2 \div 0.4$ the correlations may increase right up to 90$\%$. They become important for almost all $x$
with growing $t$ in accordance with the predicted QCD asymptotic behavior~\cite{snig, snig2}. The more comprehensive analysis of double parton distributions based on the direct numerical integration of generalized equations can be found in the quite recent paper~\cite{stirling09} (see also Ref.~\cite{sjostrand3} for estimation therein of joint interaction probability due to evolution in comparison with the factorization contribution).

\maketitle
\section{\label{sec3} Possible manifestation of correlations induced by QCD evolution}
As an effect of evolution, the double distribution functions become strongly correlated in longitudinal momentum fractions at large $Q^2$ and finite $x$ as mentioned above. On the other hand, the indications from the experimental observation of double scatterings at CDF~\cite{cdf} are not in favor of strong correlations effects in longitudinal momentum fractions. The most likely reason is that the studied kinematical domain, with relatively small $x$ values and low resolution (energy) scale, is far from the asymptotic QCD prediction. The possibility of testing double collisions at much higher resolution scales 
will open an opportunity of probing the correlations predicted by the QCD evolution 
{\it directly}. For this purpose, the double parton cross sections of the equal sign $W$ pair production (with a high resolution scale) in $pp$ collisions at 1 TeV $\leq \sqrt{s} \leq$ 14 TeV are considered in Ref.~\cite{del05}. As a main result, the contribution of the term with correlations in the equal sign $W$ pairs production might contribute almost 40$\%$ to the cross section at  $\sqrt{s}=$ 1 TeV and about 20$\%$ at 14 TeV.

\begin{figure}
\includegraphics[width=8.00cm]{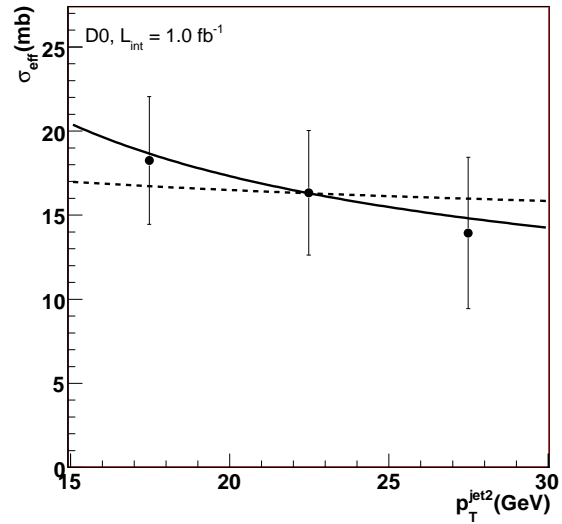}
\caption{Effective cross section $\sigma_{\rm eff}^{\rm exp}$ measured in the three $p^{\rm jet2}_T$ bins at the D0 experiment~\cite{D0}.
The solid ($k=0.5$) and dashed ($k=0.1$) lines are the results from Eq.~(\ref{prediction})
at $p_{T0}^{\rm jet2} = 22.5$ GeV and $\sigma_{\rm eff}^0=16.3$ mb. 
\label{fig:sigma_eff}}
\end{figure}

Here we would like to bring attention to the D0 measurement of the process-independent parameter of double parton scattering, $\sigma_{\rm eff}^{\rm exp}$, done as a function of the second (ordered in the transverse momentum $p_T$) jet $p_T$, $p^{\rm jet2}_T$, that can serve as a resolution scale. It is shown in Fig.~\ref{fig:sigma_eff} (see also Fig.~11 from Ref.~\cite{D0}) and can be considered as a first {\it inderect} manifestation of 
the evolution effect since $\sigma_{\rm eff}^{\rm exp}$ shows a tendency to be dependent on the resolution scale.
At first glance the dimensional parameter $\sigma_{\rm eff}$ entering into Eq.~(\ref{fourjet}) contains  the information related with the non-perturbative structure of the proton and corresponds to the overlap of the matter distributions in the colliding hadrons and therefore it is independent of the  hard process scale {\it a priori}. However the experimental effective cross section $\sigma_{\rm eff}^{\rm exp}$ is not measured directly but is calculated (extracted) using the normalization to the product of two single cross sections:
\begin{eqnarray} 
\label{dps}
\frac{\sigma_{DPS}^{\gamma+3j}}{\sigma^{\gamma j}\sigma^{jj}}= [\sigma_{\rm eff}^{\rm exp}]^{-1}
\end{eqnarray} 
in both the CDF and D0 experiments. Here $\sigma^{\gamma j}$ and $\sigma^{j j}$ are the inclusive $\gamma +$ jet and dijets cross sections, $\sigma_{DPS}^{\gamma+3j}$ is the inclusive cross section of the $\gamma + 3$ jets events produced in the double parton process. The factor 2 before $\sigma_{\rm eff}^{\rm exp}$ is not needed in this case of distinguishable scatterings unlike the four-jet process. It is worth noticing that the CDF and D0 Collaborations extract $\sigma_{\rm eff}^{\rm exp}$ without any theoretical predictions on the $\gamma +$ jet and dijets cross sections, by comparing the number of observed double parton $\gamma + 3$ jets events in one hard $p{\bar p}$ collision to the number of $\gamma + 3$ jets events with hard interactions occurring in two separate $p{\bar p}$ collisions.

At  such a normalization~(\ref{dps}) and the presence of correlations in the two-parton distributions the experimentally extracted dimensional factor $\sigma_{\rm eff}^{\rm exp}$ will be different from one $\sigma_{\rm eff}$ incoming in Eq.~(\ref{fourjet}). Indeed, in accordance with QCD evolution instead of the factorization ansatz~(\ref{factoriz}) we can write~\cite{snig03, del05}:  
\begin{eqnarray}
\label{dpdfunctions}
D_ p^{ij}(x_i,x_j,t) = D_ p^i(x_i,t)D_ p^j(x_j,t)(1-x_i-x_j)[1+\nonumber \\
R_1^{ij}(x_i,x_j,t)+R_2^{ij}(x_i,x_j,t)],
\end{eqnarray} 
where 
\begin{equation}
\label{homogeneous}
D_ p^i(x_i,t)D_ p^j(x_j,t)(1-x_i-x_j)[1+R_1^{ij}(x_i,x_j,t)]
\end{equation}
is the solution of the homogeneous generalized GLAPD evolution equation with the given initial condition $D_ p^i(x_i,0)D_ p^j(x_j,0)(1-x_i-x_j)$ which is supposed to be factorized at the reference scale $t=0 ~(Q^2=\mu^2)$ and
\begin{equation}
\label{nonhomogeneous}
D_ p^i(x_i,t)D_ p^j(x_j,t)(1-x_i-x_j)R_2^{ij}(x_i,x_j,t) 
\end{equation}
is a particular solution of the complete equation with zero initial condition
$R_2^{ij}(x_i,x_j,0)=0$. At $t>0$ this solution is always positive as it is the integral convolution of positive single distributions with the kernels of nonhomogeneous part of the evolution equation. These kernels (probabilities) are defined without negative $\delta$-function regularization and therefore are always positive unlike the kernels of homogeneous part which have negative contributions. In the kinematical range of interest for the actual case (we never exceed $x=0.1$ practically) the contribution of the term $R_1^{ij}(x_i,x_j,t)$ to Eq.~(\ref{dpdfunctions}) is negligible~\cite{del05}. Substituting 
Eq.~(\ref{dpdfunctions}) in Eq.~(\ref{fourjet}), and neglecting the factor $(1-x_i-x_j)$ as it is usually done at relatively small $x_i, x_j$, and using Eq.~(\ref{dps}) we derive
\begin{eqnarray} 
\label{relation}
[\sigma_{\rm eff}^{\rm exp}]^{-1} \simeq [\sigma_{\rm eff}]^{-1}[1+\Delta(t)],
\end{eqnarray} 
where $\Delta(t)$ is a some positive contribution induced by the correlation term
$R_2^{ij}(x_i,x_j,t)$. This contribution increases~\cite{snig03, del05} with a growth of the resolution scale $t(Q^2)$ and can be right up 40$\%$ at $Q=M_W=80.4$ GeV following Ref.~\cite{del05}. Taking into account that namely $\sigma_{\rm eff}$ is the true process-independent dimensional parameter we conclude: the experimentally extracted effective cross section $\sigma_{\rm eff}^{\rm exp}$ must decrease with the growth of the resolution scale. The experimental data in Fig.~\ref{fig:sigma_eff} show just such a tendency in spite of large experimental uncertainties which do not allow
the D0 Collaboration to make this conclusion.

Unfortunately, the calculation of correlation contribution $\Delta(t)$ to the effective cross section $\sigma_{\rm eff}^{\rm exp}$ is not yet possible in the framework of some kind of existing event generators. For instance, the double parton scatterings are implemented in the Monte-Carlo generator PYTHIA~\cite{pythia, pythia8} taking into account some correlations which, however, are not quite adequate for the case of evolution effects under consideration (unlike the theoretical investigations in Ref.~\cite{sjostrand3}). The implementation of the QCD evolution of two-parton distribution functions in some Monte-Carlo generator, as this was done for single distributions, is not a trivial task. However we can ``predict'' a functional form of 
$p_T^{\rm jet2}$-dependence of $\sigma_{\rm eff}^{\rm exp}$:
\begin{eqnarray} 
\label{prediction}
\sigma_{\rm eff}^{\rm exp} = \sigma_{\rm eff}^0[1+ k \ln(p_T^{\rm jet2}/p_{T0}^{\rm jet2}]^{-1}
\end{eqnarray} 
inspired by the explicit expression for the correlation term~\cite{snig03} and the evolution variable $t$. The results from Eq.~(\ref{prediction}) are also shown in Fig.~1 at $k=0.1$ (dashed line) and $k=0.5$ (solid line) to illustrate the two possible slopes of $\sigma_{\rm eff}^{\rm exp}$ in the observation region: visually imperceptible (practically constant) and noticeable falling slopes. The normalization point $p_{T0}^{\rm jet2} = 22.5$ GeV with $\sigma_{\rm eff}^0=16.3$ mb is fixed to reproduce the experimental value in the central $p_T^{\rm jet2}$ bin.

\maketitle
\section{\label{sec4} Conclusions}
We argue that the QCD dynamical correlations result effectively in the dependence of the experimentally {\it extracted} $\sigma_{\rm eff}^{\rm exp}$ on the resolution scale unlike naive accepted expectations. The measurements covering a larger range of the resolution scale variation with a smaller uncertainty are  needed to see the evolution effect more distinctly.

In order to investigate the more delicate  characteristics of double parton scatterings
(distributions over various kinematic variables with various kinematic cuts) it is also desirable to implement the QCD evolution of the two-parton distribution functions in some Monte Carlo event generator as this was done for the single distributions, for instance, within PYTHIA~\cite{pythia}. It is worth noticing once more that the evolution of the two-parton distribution functions has the same confidence status as the well-established evolution of the single distribution functions in the framework of the leading logarithm approximation of perturbative QCD. Therefore the experimental direct or indirect observation of this evolution effect is significant to answer to many challenging questions of yet poorly-understood aspects of QCD. 

\begin{acknowledgments}
Discussions with D.V.~Bandurin (especially), E.E.~Boos, M.N.~Dubinin, V.A.~Ilyin, V.L.~Korotkikh, L.N.~Lipatov, I.P.~Lokhtin, S.V.~Molodtsov, S.V.~Petrushanko, A.S.~Proskuryakov, V.I.~Savrin, T.~Sjostrand, P.~Skands, D.~Treleani, G.M.~Zinovjev and N.P.~Zotov are gratefully acknowledged. 
This work is partly supported by Russian Foundation for Basic Research (grants No 08-02-91001 and No 08-02-92496), Grants of President of Russian Federation for support of Leading Scientific Schools (No 1456.2008.2) and Russian Ministry for Education and Science (contract 02.740.11.0244). 
\end{acknowledgments}


\end{document}